\documentclass[10pt,fleqn]{article}
\usepackage{epsf}
\newlength{\ind}
\newlength{\indd}
\newlength{\imagewidth}
\begin{document}
\renewcommand{\d}{{{\rm d}}}
\newcommand{\eps}[2]{\centering\parbox{#1}{\epsfxsize=#1\epsfbox{#2}}}
\title{\bf Magnetic field strength from peaked synchrotron spectra}
\author{C. Hettlage and K. Mannheim\\
Universit\"{a}tssternwarte, Geismarlandstra{\ss}e 11,\\D-37083 G\"{o}ttingen, Germany}
\date{}
\maketitle
Key diagnostics of non-thermal plasmas, such as the position of the synchrotron self-absorption turnover or the Faraday rotation optical depth, depend crucially on the unobserved low-energy electron spectrum.  We investigate the effect of physically viable assumptions about the low-energy electron spectrum on the magnetic field strength as inferred from the position of the peak in the synchrotron spectrum marking the transition from optically thick to optically thin radiation. To this end the kinetic equations are written down and stationary solutions are given. In this context the inviability of the delta function approximation of the synchrotron power spectrum is discussed. Using two different approaches, we show that the field determination is independent of the assumptions about the low energy electron spectrum within an order of magnitude. As an example, we use our result to obtain a small correction to the limiting brightness temperature.
\section{Introduction}
The evaluation of the magnetic field strength of radio sources may in principle be based on the fact that their synchrotron spectra exhibit a maximum, which arises due to self absorption in the emitting plasma (Rybicki \& \mbox{Lightman}~1979). This necessitates the full knowledge of the electron\footnote{Although not stated explicitly, all results of this paper are valid for electron and electron-positron plasmas alike.} distribution. However, whereas the distribution in the optically thin regime can be inferred from the spectrum and is known to be of power law form, it remains uncertain in the optically thick regime, so that for the latter some assumption is needed. Obviously, the simplest alternative would be to assume an unbroken power law (cf. Pacholczyk~1970).

In this article, the differences between an unbroken power law and some other physically more realistic distributions are investigated. Throughout, apart from the assumption of homogeneity and isotropy, the discussion will be kept as independent as possible from specific source parameters. In Section~\ref{sec:kinetic} the kinetic equations governing emission and absorption in a synchrotron plasma are reviewed, and two simple steady state solutions are given. They motivate the use of either a power law with index -3 or a thermal electron distribution in the optically thick regime. Section~\ref{sec:magnetic} deals with the magnetic field strengths one obtains under the assumption that the transition from the optically thick to the optically thin regime corresponds to the turnover energy in the electron distribution. Piecewise power laws and thermal distributions for low energies are discussed. In Section~\ref{sec:maximum} the determination of the magnetic field strength from the maximum intensity of the spectrum is investigated. The uncertainty in the limiting brightness temperature is discussed briefly in Section~\ref{sec:temperature}. 

\section{Kinetic equations}\label{sec:kinetic}
Treating synchrotron radiation as a special case of brems\-strahl\-ung, we can see that synchrotron emission as a spontaneous process must be accompanied by the corresponding induced processes. Accordingly, it may be described by means of the Einstein coefficients, so that the Einstein relations $A_{21}=\frac{2h\nu^{3}}{c^{2}}B_{21}$ and $B_{12}=B_{21}$ may be applied. In order to simplify, in the following we assume the electron energies $\epsilon$ are highly relativistic with $\epsilon\gg\hbar\omega$ ($\omega$ being the angular frequency of an emitted photon) and that both the magnetic field and the electron distribution may be treated as isotropic. Replacing $A_{21}$ by the power spectrum $P(\epsilon,\omega)$ and employing the Einstein relations then yields the kinetic equations describing a synchrotron emitting plasma (McCray~1969, Norman \& ter Haar~1974):
\[\frac{\partial W(\omega)}{\partial t}=\int\d\epsilon\overline{P}(\epsilon,\omega)\left\{\frac{}{}f(\epsilon)+\frac{\pi^{2}c^{3}}{\omega^{2}}\epsilon^{2}\frac{\partial}{\partial\epsilon}\left(\frac{f(\epsilon)}{\epsilon^{2}}\right)W(\omega)\right\}\]
\[\frac{\partial f(\epsilon)}{\partial t}=\frac{\partial}{\partial\epsilon}\left(\int\d\omega\overline{P}(\epsilon,\omega)\left\{f(\epsilon)+\frac{\pi^{2}c^{3}}{\omega^{2}}\epsilon^{2}\frac{\partial}{\partial\epsilon}\left(\frac{f(\epsilon)}{\epsilon^{2}}\right)W(\omega)\right\}\right)\]
In these formulae, $f$ stands for the electron density per energy $\epsilon$, $W$ for the photon energy density per angular frequency $\omega$. $\overline{P}$ is the angular average of $P$, i.e. $\overline{P}=\int_{4\pi}P\d\Omega$. The kinetic equations obtained in this manner remain valid in case of a homogeneous magnetic field and an isotropic photon distribution. In the stationary case comparison of the kinetic equations and the transfer equation $\frac{\d I_{\nu}}{\d s}=j_{\nu}-\alpha_{\nu}I_{\nu}$ with emission coefficient $j_{\nu}$, absorption coefficient $\alpha_{\nu}$, and path length $s$ yields an expression for $\alpha_{\nu}$:

\begin{equation}
\alpha_{\nu}=-\int\d\epsilon\overline{P}(\epsilon,\omega)\frac{\pi^{2}c^{2}}{\omega^{2}}\epsilon^{2}\frac{\partial}{\partial\epsilon}\left(\frac{f(\epsilon)}{\epsilon^{2}}\right)
\label{eq:alpha}
\end{equation}
Turning now from these rather general considerations to the specific example of synchrotron radiation, one starts by defining $x\equiv\omega/\omega_{\rm c}$ and $F(x)\equiv x\int_{x}^{\infty}K_{5/3}(\xi)\d\xi$, where $\omega_{\rm c}\equiv3\epsilon^{2}eB\sin\alpha/2m_{\rm e}^{3}c^{5}$ with the elementary charge $e$, the electron mass $m_{\rm e}$ and the pitch angle $\alpha$ between magnetic field and electron velocity. $K_{5/3}$ constitutes the modified Bessel function of order 5/3. Using these definitions, one may express the synchrotron power spectrum as $P(\epsilon,\omega)=(\sqrt{3}e^{3}B\sin\alpha/2\pi m_{\rm e}c^{2})^{1/2}F(x)$ (Ginzburg \& \mbox{Syrovatskii}~1965). Its angular average is given by $\overline{P}(\epsilon,\omega)=Q(\tilde{x})\left.P(\epsilon,\omega)\right|_{\sin\alpha=1}$ with $\tilde{x}\equiv\left.x\right|_{\sin\alpha=1}$, where for
\[Q(\tilde{x})=\frac{1}{2F(\tilde{x})}\int_{-1}^{+1}\d\mu\sqrt{1-\mu^{2}}F\left(\frac{\tilde{x}}{\sqrt{1-\mu^{2}}}\right)\]
the approximation $Q(\tilde{x})\approx\sum_{n=0}^{3}Q_{n}\tilde{x}^{n}\equiv0.822533-0.181121\,\tilde{x}+0.0370118\,\tilde{x}^{2}-0.00316623\,\tilde{x}^{3}$ can be used. (Because of $\overline{P}(\epsilon,\omega)\leq\left.P(\epsilon,\omega)\right|_{\sin\alpha=1}$, $Q(\tilde{x})\leq1$, so that for $\tilde{x}\longrightarrow\infty$ the approximation becomes arbitrarily inaccurate. However, $Q(\tilde{x})$ will always be part of an integrand containing ${\rm e}^{-\tilde{x}}$, which renders this inaccuracy irrelevant.) Concerning stationary distributions, two important solutions of the kinetic equations can be stated (Norman \& ter Haar~1974). Firstly, the distributions may be those of thermal equilibrium in the Rayleigh-Jeans limit, i.e. $f(\epsilon)=f_{0}\exp\left(-\epsilon/\pi^{2}c^{3}\Omega_{0}\right)$, $W(\omega)=\Omega_{0}\omega^{2}$, where $\phi_{0}$ and $\Omega_{0}$ are arbitrary positive constants. Secondly, the only viable power law for the electron distribution has the index -3, so that $f(\epsilon)=f_{0}\epsilon^{-3}$. The corresponding photon distribution has the form $W(\omega)=\frac{1}{5\pi^{2}}\sqrt{\frac{2m_{\rm e}^{3}}{3ceB}}\omega^{5/2}$. Finally, one may note that the form of the kinetic equations precludes the use of the delta approximation $\overline{P}(\epsilon,\omega)=C_{1}\epsilon^{2}\delta(\omega-C_{2}\epsilon^{2})$ ($C_{1}$, $C_{2}$ being constants) for the power spectrum, as, despite of yielding the correct photon distributions, this approximation allows any electron distribution, which contradicts the fact that only one power law index is allowed.

\section{Magnetic field strength}\label{sec:magnetic}
\begin{figure}
\eps{\imagewidth}{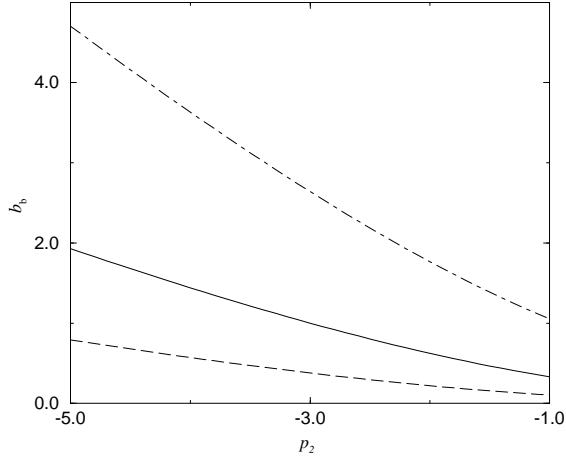}
\caption{Ratio $b_{\rm b}$ of magnetic field strengths due to $f_{\rm b}$ and $f_{\rm a}$ as a function of the power law index $p_{2}$ for $N=1$ (solid), 0.5 (long dashed) and 2 (dot-dashed line).}\label{fig:modelb}
\end{figure}

\begin{figure}
\eps{\imagewidth}{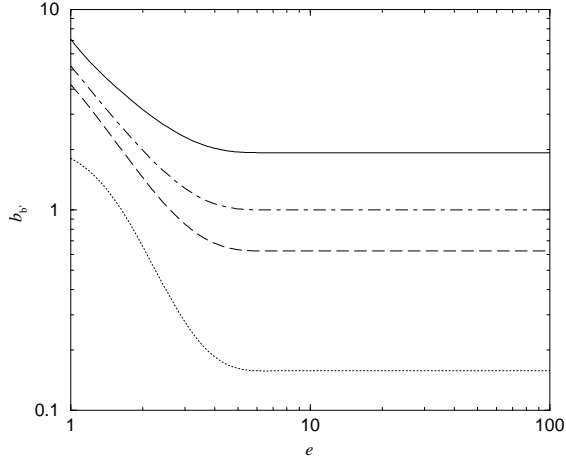}
\caption{Ratio $b_{{\rm b}^{\prime}}$ of magnetic field strengths due to $f_{{\rm b}^{\prime}}$ and $f_{\rm a}$ as a function of the ratio $e$ of turnover and cutoff energy of $f_{{\rm b}^{\prime}}$. $N=1$ is assumed, and the lines are given for the power law indexes $p_{2}=-5$ (solid), -3 (dot-dashed), -2 (long dashed), and 0 (dotted line).}\label{fig:modelbprime}
\end{figure}

\begin{figure}
\eps{\imagewidth}{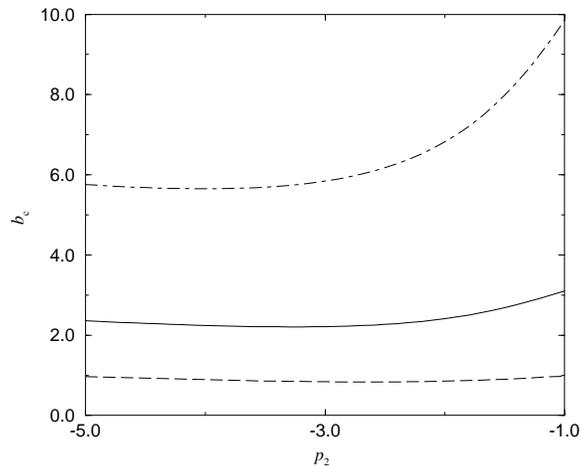}
\caption{Ratio $b_{\rm c}$ of magnetic field strengths due to $f_{\rm c}$ and $f_{\rm a}$ as a function of the power law index $p_{2}$ for $N=1$ (solid), 0.5 (long dashed) and 2 (dot-dashed line).}\label{fig:modelc}
\end{figure}

A typical synchrotron spectrum contains a turnover, marking the transition from optically thick to optically thin emission. Whereas, from observation, the optically thin emission (and thus the electron spectrum for sufficiently high energies) can generally be taken to be a power law with some index $p_{2}$, the precise form of the optically thick spectrum is hard to detect. Therefore the electron spectrum is taken to be of the form
\[f(\epsilon)=\left\{\begin{array}{lcl}
                        F(\epsilon) & , & \epsilon<\epsilon_{\rm T}\\
                        f_{0}\epsilon^{p_{2}} & , & \epsilon\geq\epsilon_{\rm T}
                     \end{array}\right.\]
where various functions $F$ will be discussed. The simplest choice is to assume an unbroken power law spectrum, i.e. $F(\epsilon)=f_{0}\epsilon^{p_{2}}$. Quantities relating to this choice are denoted by the index a. Inserting $f_{\rm a}(\epsilon)$ and the (averaged) synchrotron power spectrum $\overline{P}(\epsilon,\omega)$ into equation~(\ref{eq:alpha}), one obtains, using the relation $\int x^{\mu}F(x)\d x=\frac{2^{\mu+1}}{\mu+2}\Gamma\left(\frac{\mu}{2}+\frac{7}{3}\right)\Gamma\left(\frac{\mu}{2}+\frac{2}{3}\right)$ with $\mu>-\frac{4}{3}$ (Westfold~1959),
\begin{equation}
\alpha_{\nu}=\sum_{n=0}^{3}\alpha_{0}f_{0}R_{\rm a}^{-\frac{p_{2}}{2}}\nu^{\frac{p_{2}}{2}-2}h(p_{2},n),
\label{eq:alphaa}
\end{equation}
where $\alpha_{0}\equiv\frac{\sqrt{3}e^{3}B}{8\pi m_{\rm e}}$, $R\equiv\frac{3eB}{4\pi m_{\rm e}^{3}c^{5}}$ and
\[h(p,n)\equiv Q_{n}\left(1-\frac{p}{2}\right)\frac{2^{n-\frac{p}{2}}}{n+1-\frac{p}{2}}\Gamma\left(\xi_{1}\right)\Gamma\left(\xi_{2}\right),\]
where the $\xi_{i}$ are defined as $\xi_{1}\equiv(6n+22-3p)/12$ and $\xi_{2}\equiv(6n+2-3p)/12$. Note that equation~(\ref{eq:alphaa}) can easily be solved for the magnetic field strength $B$. A second, slightly more general choice for $f$, which will be indicated by the index b, is a piecewise power law, so that $F(\epsilon)=f_{0}\epsilon_{\rm T}^{p_{2}-p_{1}}\epsilon^{p_{1}}$. Here the constant $\epsilon_{\rm T}^{p_{2}-p_{1}}$ is implied by demanding continuity in $\epsilon_{\rm T}$. Using the approximation $F(x)\approx F_{0}\tilde{x}^{\phi}{\rm e}^{-\tilde{x}}=1.8\tilde{x}^{0.3}{\rm e}^{-\tilde{x}}$ (Melrose~1980) together with $\tilde{x}(\epsilon,\nu)\equiv\left.x\right|_{\sin\alpha=1}=\nu/R\epsilon^{2}$ and
\[g(p,n,\tilde{x}_{1},\tilde{x}_{2})\equiv-F_{0}Q_{n}\left\{\zeta_{3}\Gamma_{\tilde{x}_{1}}^{\tilde{x}_{2}}\left(\zeta_{1}\right)-\Gamma_{\tilde{x}_{1}}^{\tilde{x}_{2}}\left(\zeta_{2}\right)\right\}\]
(where $\zeta_{1}\equiv n+\phi-p/2$, $\zeta_{2}\equiv n+\phi-p/2+1$ and $\zeta_{3}\equiv n+\phi-1$) one gets for the ratio $\alpha_{\nu_{1},{\rm b}}/\alpha_{\nu_{2},\rm a}$ of the absorption coefficients computed with $f_{\rm a}$ and $f_{\rm b}$
\begin{equation}
\frac{\alpha_{\nu_{1},{\rm b}}}{\alpha_{\nu_{2},{\rm a}}}=\frac{\sum_{n=0}^{3}\left\{\alpha_{\rm den}(\nu_{1},\epsilon_{\rm T,b},p_{1},p_{2},n)\right\}}{\sum_{n=0}^{3}\alpha_{0,\rm a}f_{0}R_{\rm a}^{-\frac{p_{2}}{2}}\nu_{2}^{\frac{p_{2}}{2}-2}h(p_{2},n)}
\label{eq:alpharatio}
\end{equation}
where
\settowidth{\ind}{+}
\settowidth{\indd}{=}
\begin{eqnarray*}
\lefteqn{\alpha_{\rm den}(\nu_{1},\epsilon_{\rm T,b},p_{1},p_{2},n)}\\
& & =\hspace*{\ind}\alpha_{0,{\rm b}}f_{0}R_{\rm b}^{-\frac{p_{2}}{2}}\nu_{1}^{\frac{p_{2}}{2}-2}g(p_{2},n,0,\tilde{x}(\nu_{1},\epsilon_{\rm T,b}))\\
& & \hspace*{\indd}+\mbox{}\alpha_{0,\rm b}f_{0}\epsilon^{p_{2}-p_{1}}R_{\rm b}^{-\frac{p_{1}}{2}}\nu_{1}^{\frac{p_{1}}{2}-2}g(p_{1},n,\tilde{x}(\nu_{1},\epsilon_{\rm T,b}),\infty).
\end{eqnarray*}
Now the synchrotron power spectrum takes its maximum at the frequency $\nu_{\rm mp}(\epsilon)=0.29\frac{\pi}{4}R\epsilon^{2}$, so that $\tilde{x}(\nu_{\rm mp}(\epsilon_{\rm T}),\epsilon_{\rm T})=0.29\frac{\pi}{4}R\epsilon_{\rm T}^{2}/R\epsilon_{\rm T}^{2}=0.29\frac{\pi}{4}$. As $\epsilon_{\rm T}$ constitutes the transition between the parts of the electron spectrum yielding optically thin and optically thick radiation, it is reasonable to assume that the optical depth corresponding to $\nu_{\rm mp}(\epsilon_{\rm T})$ is approximately 1, i.e. that for a homogeneous plasma $\alpha_{\nu_{\rm mp}(\epsilon_{\rm T})}L\approx1$, where $L$ denotes the spatial extension of the plasma. (In the case of an unbroken power law, this condition is used to define $\epsilon_{\rm T}$.) Hence $1\approx\alpha_{\nu_{\rm mp}(\epsilon_{\rm T}),{\rm b}}/\alpha_{\nu_{\rm mp}(\epsilon_{\rm T}),{\rm a}}$. If one allows for different turnover energies, this may be generalized to $1\approx\alpha_{\nu_{\rm mp}(\epsilon_{\rm T,b}),{\rm b}}/\alpha_{\nu_{\rm mp}(\epsilon_{\rm T,a}),{\rm a}}$, and using equation~(\ref{eq:alpharatio}) and $\tilde{x}(\nu_{\rm mp}(\epsilon_{\rm T}),\epsilon_{\rm T})=0.29\frac{\pi}{4}$ together with the ratio $r_{\rm b}\equiv\epsilon_{\rm T,b}/\epsilon_{\rm T,a}$ one can obtain the ratio $b_{\rm b}=B_{\rm b}/B_{\rm a}$ of the magnetic field strengths for $f_{\rm a}$ and $f_{\rm b}$, respectively:
\[b_{\rm b}=r_{\rm b}^{p_{2}-4}\cdot\frac{\sum_{n=0}^{3}\left\{\gamma_{1}+\left(\frac{0.29\pi}{4}\right)^{\frac{1}{2}(p_{1}-p_{2})}\gamma_{2}\right\}}{\sum_{n=0}^{3}h(p_{2},n)}\]
Here the $\gamma_{i}$ denote $\gamma_{1}=g(p_{2},n,0,\frac{0.29}{4}\pi)$ and $\gamma_{2}=g(p_{1},n,\frac{0.29}{4}\pi,\infty)$. If the turnover frequency is independent of the electron spectrum (i.e. if $\nu_{\rm mp,a}(\epsilon_{\rm T,a})=\nu_{\rm mp,b}(\epsilon_{\rm T,b})$), the relation $br_{\rm b}^{2}=1$ holds valid. Hence we assume that $br_{\rm b}^{2}=N$ with some arbitrary, but constant $N$. In view of the discussion at the end of the previous section, one should take $p_{1}=-3$. For $N$=0.25, 1, and 2, the plots of $b_{\rm b}$ versus $p_{2}$ corresponding to this choice of $p_{1}$ are shown in Fig.~\ref{fig:modelb}. Note that $b_{\rm b}(p_{2}=-3)\neq1$ for $N=1$ results from the use of the approximation $F(x)\approx1.8x^{0.3}e^{-x}$ in computing $\alpha_{\nu,{\rm b}}$. A simple generalization of $f_{\rm b}$ is achieved by dropping the assumption that the electron spectrum extends down to zero energy (which, due to the finite electron rest energy, is questionable anyway), i.e. by taking a spectrum $f_{{\rm b}^\prime}(\epsilon)\equiv f_{\rm b}(\epsilon)\Theta(\epsilon-\epsilon_{0})$ rather than $f_{\rm b}(\epsilon)$, where $\epsilon_{0}>0$ and $\Theta$ denotes the Heaviside function. This merely involves replacing $\tilde{x}_{2}=\infty$ in the argument of $g(p,n,\tilde{x}_{1},\tilde{x}_{2}=\infty)$ by $\tilde{x}(\nu,\epsilon_{0})$ with the cutoff energy $\epsilon_{0}$. If one introduces the ratio $e\equiv\epsilon_{\rm T,b}/\epsilon_{0}$, one obtains $\tilde{x}(\nu_{\rm mp}(\epsilon_{\rm T,b}),\epsilon_{0})=0.29\frac{\pi}{4}e^{2}$ and thus for the ratio $b_{{\rm b}^{\prime}}=B_{{\rm b}^{\prime}}/B_{\rm a}$
\[b_{{\rm b}^{\prime}}=r_{\rm b}^{p_{2}-4}\cdot\frac{\sum_{n=0}^{3}\left\{\gamma_{1}+\left(\frac{0.29\pi}{4}\right)^{\frac{1}{2}(p_{1}-p_{2})}\bar{\gamma}_{2}\right\}}{\sum_{n=0}^{3}h(p_{2},n)},\]
where $\bar{\gamma}_{2}\equiv g(p_{1},n,\frac{0.29}{4}\pi,0.29\frac{\pi}{4}e^{2})$ and $\gamma_{1}$ is defined as above. Again, $p_{1}$ should be set to -3. The plots of $b_{{\rm b}^{\prime}}$ as a function of $e$ resulting from this formula are given in Fig.~\ref{fig:modelbprime} for several values of $p_{2}$ and $N=1$. As expected, in the limit of $e\longrightarrow\infty$ (no energy cutoff) $b_{{\rm b}^{\prime}}$ approaches $b_{\rm b}$; for values of $e$ exceeding 10, the two are essentially indistinguishable. Referring to the examples in the previous section, it is tempting to assume, as a further possibility for the electron spectrum, a thermal distribution for low energies, i.e. to choose $F(\epsilon)=f_{0}{\rm e}^{\beta\epsilon_{\rm T}}\epsilon_{\rm T}^{p_{2}-2}\epsilon^{2}{\rm e}^{-\beta\epsilon}$, where $\beta\equiv1/k_{\rm B}T$ with the Boltzmann constant $k_{\rm B}$ and the temperature $T$. One can see easily that continuity in $\epsilon_{\rm T}$ is ensured. The quantities relating to this choice of $f(\epsilon)$ are marked by the index c. Inserting $f_{\rm c}(\epsilon)$ into equation~(\ref{eq:alpha}) and using $\tilde{x}(\nu_{\rm mp}(\epsilon_{\rm T,c}),\epsilon_{\rm T,c})=0.29\frac{\pi}{4}$ yields
\begin{eqnarray*}
\lefteqn{\alpha_{\nu,{\rm c}}(\bar{\nu})=\sum_{n=0}^{3}\left\{\alpha_{0,c}f_{0}R_{\rm c}^{-\frac{p_{2}}{2}}\bar{\nu}^{\frac{p_{2}}{2}-2}\gamma_{1}\right.}\\
& & \hspace*{4.3em}\left.\mbox{}+\alpha_{0,{\rm c}}f_{0}\Delta(p_{1},p_{2})\beta{\rm e}^{\beta\epsilon_{\rm T,c}}\tilde{g}(\beta\epsilon_{\rm T,c},n)\right\},
\label{eq:alphac}
\end{eqnarray*}
where $\Delta(p_{1},p_{2})\equiv\left(\frac{0.29\pi}{4}\right)^{-\frac{p_{2}}{2}+1}R_{\rm c}^{-\frac{p_{2}}{2}-\frac{1}{2}}\bar{\nu}^{\frac{p_{1}}{2}-\frac{3}{2}}$, $\bar{\nu}\equiv\nu_{\rm mp}(\epsilon_{\rm T,c})$, and
\[\tilde{g}(u,n)=\frac{1}{2}F_{0}Q_{n}\int_{\frac{0.29}{4}\pi}^{\infty}\d\tilde{x}\tilde{x}^{n+\phi-\frac{5}{2}}{\rm e}^{-\tilde{x}}{\rm e}^{-\sqrt{\frac{0.29}{4}\pi}u\frac{1}{\sqrt{x}}}.\]
Now in order to simplify the discussion, we assume $N_{\rm th}$ to be minimized. This assumption is motivated by the low photon density in the optically thin part of jet spectra, as implied by polarisation measurements (Jones~1988). Hence one demands $\frac{\d N_{\rm th}}{\d\epsilon_{\rm T,c}}=0$, which together with $u\equiv\beta\epsilon_{\rm T,c}$ leads to
\begin{eqnarray*}
0=2(p_{2}-2)({\rm e}^{u}-1)u^{-3}+2({\rm e}^{u}-(p_{2}-1))u^{-2}-p_{2}u^{-1}.
\end{eqnarray*}
Let $u_{\rm min}=u_{\rm min}(p_{2})$ be the solution of this implicit equation.
Then using Eqs.~\ref{eq:alphaa} and~\ref{eq:alphac} and defining $\tilde{\gamma}\equiv\tilde{g}(u_{\rm min},n)$, one obtains for the ratio $b_{\rm c}\equiv B_{\rm c}/B_{\rm a}$
\[b_{\rm c}=r_{\rm c}^{p_{2}-4}\cdot\frac{\sum_{n=0}^{3}\left\{\gamma_{1}+\left(\frac{0.29\pi}{4}\right)^{-\frac{p_{2}}{2}+\frac{3}{2}}u_{\rm min}{\rm e}^{u_{\rm min}}\tilde{\gamma}\right\}}{\sum_{n=0}^{3}h(p_{2},n)},\]
where $r_{\rm c}=\epsilon_{\rm T,c}/\epsilon_{\rm T,a}$, the analogue of $r_{\rm b}$, is assumed to be given by $br_{\rm c}^{2}=N$ with some constant $N$. The plots of $b_{\rm c}$ versus $p_{2}$ for several values of $N$ are shown in Fig.~\ref{fig:modelc}. Note that the restriction to values of $p_{2}$ less than -1 is necessitated by the demand that $u_{\rm min}$ be positive.

\section{Maximum intensity}\label{sec:maximum}
So far, it has been assumed that the optical depth of the frequency corresponding to the turnover energy is equal to 1. However, from an observational point of view, it is clearly more desirable to deduce the magnetic field from the position of the maximum intensity. In order to achieve this aim, one assumes the radiative transfer to be one-dimensional and starts from the general formula for the intensity $I_{\nu}$ of a homogeneous source (Pacholczyk~1970),
\begin{equation}
I_{\nu}=\frac{j_{\nu}}{\alpha_{\nu}}(1-{\rm e}^{-\alpha_{\nu}L}). \label{eq:intensity}
\end{equation}
Now for a piecewise power law with a lower cutoff energy $e^{-1}\epsilon_{\rm T}$, using the notations introduced in the previous section, one may write the absorption coefficient as $\alpha_{\nu}=\sum_{i=1}^{2}\sum_{n=0}^{3}\alpha_{0}A_{i}R^{-\frac{p_{i}}{2}}\nu^{\frac{p_{i}}{2}-2}g(p_{i},n,\tilde{x}_{1i},\tilde{x}_{2i})$, where $\tilde{x}_{11}=\tilde{x}(\nu,e^{-1}\epsilon_{\rm T})$, $\tilde{x}_{12}=\tilde{x}_{21}=\tilde{x}(\nu,\epsilon_{\rm T})$, $\tilde{x}_{22}=\tilde{x}(\nu,\infty)=0$, $A_{1}=f_{0}\epsilon_{\rm T}^{p_{2}-p_{1}}$, and $A_{2}=f_{0}$. If one introduces the turnover frequency $\nu_{0}\equiv\nu_{\rm mp}(\epsilon_{\rm T})$, the $\tilde{x}_{ik}$ become functions of this frequency only, i.e. they may be considered to be independent of the magnetic field. Accordingly, $\alpha_{\nu}$ takes the form $\alpha_{\nu,\nu_{0}}=\sum_{i=1}^{2}f_{0}F_{\alpha i}(\nu,\nu_{0})G_{\alpha i}(B)$ with polynomial functions $G_{\alpha i}(B)=B^{\kappa_{i}}$. However, from $\alpha_{0}\propto R\propto B$ and $A_{1}\propto\epsilon_{\rm T}^{p_{2}-p_{1}}\propto\nu_{0}^{1/2}B^{(p_{2}-p_{1})/2}$ it follows that $\kappa_{1}=\kappa_{2}=1-\frac{p_{2}}{2}\equiv\kappa$, yielding $\alpha_{\nu}=G_{\alpha}(\nu,\nu_{0})B^{\kappa}$ and hence $\frac{\partial}{\partial\nu}\alpha_{\nu}=f_{0}F_{\partial\alpha}(\nu,\nu_{0})B^{\kappa}$. Similarly one obtains $j_{\nu}=f_{0}F_{j}(\nu,\nu_{0})B^{\lambda}$ and $\frac{\partial}{\partial\nu}j_{\nu}=f_{0}F_{\partial j}(\nu,\nu_{0})B^{\lambda}$ (with $\lambda=\frac{1}{2}(1-p_{2}$)) for the emissivity and its derivative.

\begin{figure}
\eps{\imagewidth}{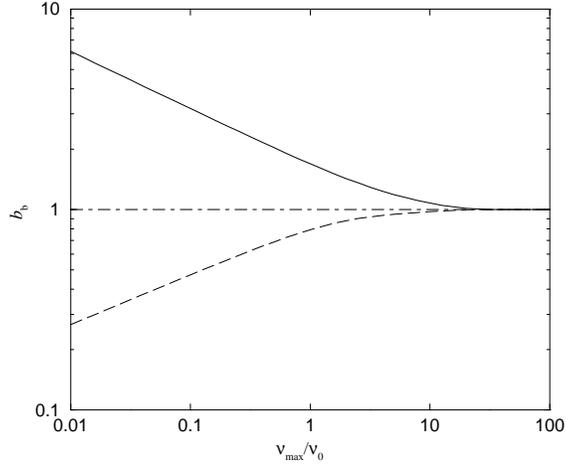}
\caption{The ratio $b_{\rm b}$ as a function of $\nu_{\rm max}/{\nu_{0}}$ for several power law indexes $p_{2}$. The lines correspond to $p_{2}=-5$ (solid), -3 (dot-dashed), and -2 (long dashed line).}\label{fig:enu0}
\end{figure}

\begin{figure}
\eps{\imagewidth}{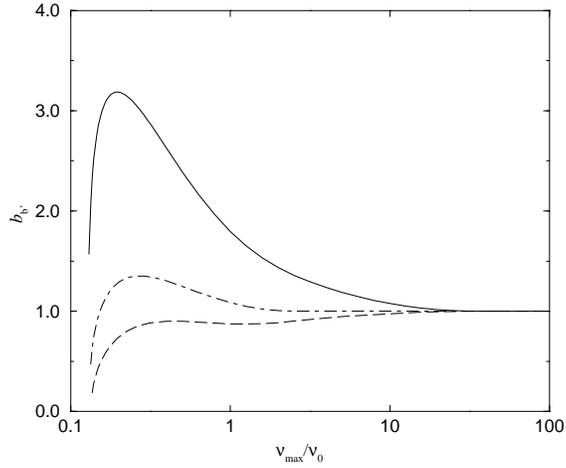}
\caption{The ratio $b_{\rm b^{\prime}}$ for $e=4$ as a function of $\nu_{\rm max}/{\nu_{0}}$ for the same power law indexes as in fig~\ref{fig:enu0}.}\label{fig:enu025}
\end{figure}

\begin{figure}
\eps{\imagewidth}{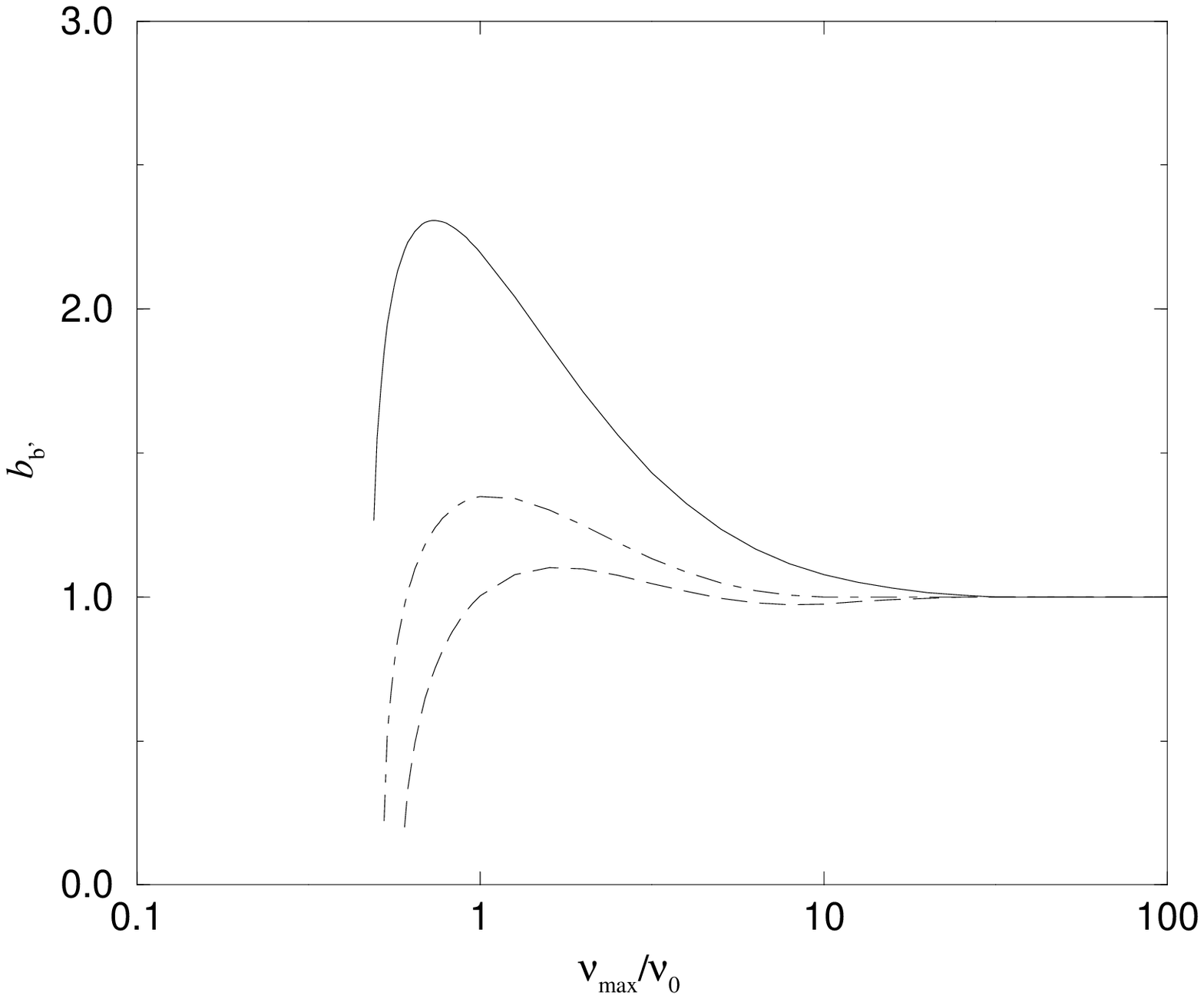}
\caption{The same as Fig.~\ref{fig:enu025}, but for $e=2$.}\label{fig:enu05}
\end{figure}

The intensity takes its maximum when its derivative vanishes. Therefore, by differentiating equation~(\ref{eq:intensity}) one obtains the condition
\begin{eqnarray}
\lefteqn{0=C_{1}(\nu_{\rm max},\nu_{0})(1-{\rm e}^{-C_{2}(\nu_{\rm max},\nu_{0})x})}\nonumber\\
& & \hspace*{4.5em}\mbox{}+C_{3}((\nu_{\rm max},\nu_{0}))x{\rm e}^{-C_{2}(\nu_{\rm max},\nu_{0})x}, \label{eq:xmax}
\end{eqnarray}
where $x\equiv f_{0}LB^{\kappa}$, $C_{1}=F_{\partial j}/F_{\alpha}-(F_{j}/F_{\alpha}^{2})F_{\partial\alpha}$, $C_{2}=F_{\alpha}$, and $C_{3}=(F_{j}/F_{\alpha})F_{\partial\alpha}$. Note that this equation is formally independent of both the source extension and the overall electron density (i.e. of $f_{0}$). Only when computing the magnetic field strength by means of $B=(x/L)^{1/\kappa}$ these parameters reenter. Clearly, for two models with the same values of $Lf_{0}$, $\nu_{\rm max}$, and $\nu_{0}$, the ratio of the magnetic field strengths to the power of $\kappa$ equals that of the $x$ obtained from equation~(\ref{eq:xmax}). Hence, it is straightforward to use the ratios $b_{\rm b}$ and $b_{\rm b^{\prime}}$ as defined in the previous section. Incidentally, this leads to a further simplification: Due to the fact that $\tilde{x}(\nu,e^{-1}\epsilon_{\rm T})$ is a function of $y\equiv\nu/\nu_{0}$ and that $A_{1}\propto\epsilon_{\rm T}^{(p_{2}-p_{1})}\propto\nu_{0}^{\frac{1}{2}(p_{2}-p_{1})}$, the functions $G$ defined above may be written as $G_{\alpha}(\nu,\nu_{0})=\nu_{0}^{\frac{p_{2}}{2}-2}\bar{G}_{\alpha}(y)$, $G_{\partial\alpha}(\nu,\nu_{0})=\nu_{0}^{\frac{p_{2}}{2}-3}\bar{G}_{\partial\alpha}(y)$, $G_{j}(\nu,\nu_{0})=\nu_{0}^{\frac{p_{2}}{2}+\frac{1}{2}}\bar{G}_{j}(y)$, and $G_{\partial j}(\nu,\nu_{0})=\nu_{0}^{\frac{p_{2}}{2}-\frac{1}{2}}\bar{G}_{\partial j}(y)$, which implies $C_{1}(\nu,\nu_{0})=\nu_{0}^{\frac{3}{2}}\bar{C}_{1}(y)$, $C_{2}(\nu,\nu_{0})=\nu_{0}^{\frac{p_{2}}{2}-2}\bar{C}_{2}(y)$, and $C_{3}(\nu,\nu_{0})=\nu_{0}^{\frac{p_{2}}{2}-\frac{1}{2}}\bar{C}(y)$. Defining $\bar{x}(y)$ as the solution of
\[\bar{C}_{1}(y)(1-{\rm e}^{-\bar{C}_{2}(y)\bar{x}(y)})+\bar{C}_{3}(y)\bar{x}(y){\rm e}^{-\bar{C}_{2}(y)\bar{x}(y)}=0\]
and multiplying both sides of this equation by $\nu_{0}^{3/2}$, one can see easily that equation~(\ref{eq:xmax}) is solved by $x(\nu_{\rm max},\nu_{0})=\nu_{0}^{-\frac{p_{2}}{2}+2}\bar{x}(y_{\rm max})$, where $y_{\rm max}\equiv\nu_{\rm max}/\nu_{0}$. Accordingly, the ratios $x_{\rm b}/x_{\rm a}$ and $x_{\rm b^{\prime}}/x_{\rm a}$ of the $x$ computed for the electron distributions a, b and ${\rm b}^{\prime}$ of the previous section depend on $y_{\rm max}$ only. However, (assuming $L$ and $f_{0}$ are the same for all distributions) these ratios clearly constitute the ratios of the respective magnetic field strengths to the power of $\kappa$, so that $b_{\rm b}\equiv b_{\rm b}(y_{\rm max})=(x_{\rm b}/x_{\rm a})^{1/\kappa}$, $b_{\rm b^{\prime}}\equiv b_{\rm b^{\prime}}(y_{\rm max})=(x_{\rm b^{\prime}}/x_{\rm a})^{1/\kappa}$. Hence the magnetic field ratios can be computed by means of solving equation~(\ref{eq:xmax}) for the distributions involved. For several values of $p_{2}$ and $e$ the results obtained numerically in this way are plotted in Figs.~\ref{fig:enu0} to~\ref{fig:enu05}. 

\begin{figure}
\eps{\imagewidth}{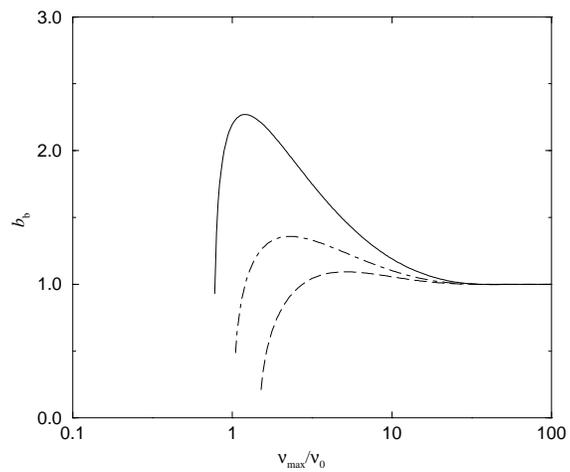}
\caption{$b_{\rm b}$ as a function of $\nu_{\rm max}/\nu_{\rm 0}$ for $p_{1}=2$ and the same values of $p_{2}$ as in fig~\ref{fig:enu0}.}\label{fig:penu0}
\end{figure}

Now although this choice is not motivated by the discussion of Section~\ref{sec:kinetic}, it is instructive to examine an electron distribution with a positive power law index $p_{1}$ in the low energy regime. For the sake of definiteness, $p_{1}=2$ is taken as an example, which may be regarded as a crude approximation of the thermal distribution. One then obtains the ratios $b_{\rm b}$ plotted in Fig.~\ref{fig:penu0}. Clearly, the result resembles that for $p_{1}=-3$ and $e=2$, which implies that the contribution of the low energy part of the electron distribution is rather small.

$x_{\rm a}$ and thus $B_{\rm a}$ do not depend on $\nu_{0}$ (see equation~(\ref{eq:xa}) below). Hence, assuming some specific value for $\nu_{\rm max}$, the ratios $b_{i}$ are given by $b_{i}(\nu_{\rm max}/\nu_{0})=CB_{i}(\nu_{\rm max}/\nu_{0})$, where the constant $C$ is independent of the spectrum~$i$. Therefore $B_{i}(z_{1})/B_{j}(z_{2})=b_{i}(z_{1})/b_{j}(z_{2})$ (where $z\equiv\nu_{\rm max}/\nu_{0})$, which allows a direct comparison of Figs.~\ref{fig:enu0} to~\ref{fig:penu0}. This clearly shows that the magnetic fields obtained for the various electron spectra discussed in these figures are of the same order of magnitude.\footnote{Obviously, this is not true, if, in Figs.~\ref{fig:enu025} to~\ref{fig:penu0}, $\nu_{\rm max}/\nu_{0}$ happens to be smaller than the value corresponding to the maximum of the relevant curve. In that case, however, due to the steepness of the curve, one should distrust the magnetic field determination anyway.}

Obviously, in order to get $x_{\rm b}$ (or $x_{\rm b^{\prime}}$) instead of the ratio $x_{\rm b}/x_{\rm a}$ (or $x_{\rm b^{\prime}}/x_{\rm a}$, respectively), one needs to know the value of $x_{\rm a}$. Now, as is evident from the above discussion, $x_{\rm a}$ must be of the form $x_{\rm a}(\nu_{\rm max},\nu_{0})=\nu_{0}^{-\frac{p_{2}}{2}+2}\bar{x}(y_{\rm max})$. However, concerning an unbroken power law, the turnover frequency has no physical meaning, as there is no turnover at all. Still, it may be retained as a formal parameter, so that the distribution is viewed as a piecewise power law, where the power happens to be the same for all pieces. This implies that for $\nu_{0}$ any value may be assumed. Choosing $\nu_{0}=\nu_{\rm max}$, one obtains $y_{\rm max}=1$ and thus
\begin{equation}
x_{\rm a}=\bar{x}(1)\nu_{\rm max}^{-\frac{p_{2}}{2}+2}.\label{eq:xa}
\end{equation}
Finally, an important technical remark should be made: The aim of the foregoing discussion was to gain results depending on as few parameters as possible. In order to achieve this, the turnover frequency rather than energy was employed. However, in a given model, the latter rather than the former will be given. In this case the magnetic field ratio cannot simply be read off, and the field strength has to be inferred from demanding self-consistency: Assuming some specific values of $L$, $f_{0}$ and the magnetic field $B$, one can compute the turnover frequency from the corresponding energy and thus obtain the ratio $x_{\rm b}/x_{\rm a}$ (or $x_{\rm b^{\prime}}/x_{\rm a}$). Together with the value of $x_{\rm a}$ taken from equation~(\ref{eq:xa}) one gets the value of $x_{\rm b}$ (or $x_{\rm b^{\prime}}$), leading to the magnetic field strength, which obviously should equal the one originally assumed.

\section{An example: peak brightness temperature}\label{sec:temperature}
As a straightforward example of using the results obtained so far, we discuss the maximum brightness temperature of a (synchrotron) source with an isotropic magnetic field (cf. Kellermann~1974). To do so, one starts by noting that the ratio of the synchrotron and (to first order) inverse Compton losses is given by
\[\frac{P_{\rm c}}{P_{\rm syn}}=\frac{u_{\rm rad}}{u_{B}}=\frac{L/(\pi\rho^{2}c)}{B^{2}/(8\pi)}=\frac{32\pi\xi_{\rm geom}R^{2}F_{\rm max}\nu_{\rm c}}{\rho^{2}c B^{2}}.\]
Here, $u_{\rm rad}$ and $u_{B}$ denote the energy density of the radiation and the magnetic field, respectively, $\rho$ the radius of the source, $R$ its distance from the observer, and $\nu_{\rm c}$ the upper cut-off frequency. Clearly, $\rho=R\theta/2$ with the angular size $\theta$. The luminosity $L$ is approximated as $L\approx\xi_{\rm geom}4\pi R^{2}F_{\rm m}\nu_{\rm c}$, where $F_{\rm m}$ is the observed maximum flux density and $\xi_{\rm geom}$ takes into account any deviation from isotropic radiation.

The solid angle of the source (as seen from the observer) is given by $\pi\theta^2/4$. Hence, assuming that the one-dimensional transfer equation (equation~(\ref{eq:intensity})) may be applied, one can show from the formulae for the absorption and emission coefficient that for an unbroken power law electron distribution with index $p$ the magnetic field strength may be expressed as
\begin{eqnarray*}
\lefteqn{B_{\rm pl}=2.488\times10^{-62}\left(\frac{a(p)}{b(p)}\right)^{2}(1-{\rm e}^{-\tau_{\rm max}})\cdot}\\
& & \hspace*{7.21em}\mbox{}\cdot\left(\frac{4F_{\rm max}/(\pi\theta^2)}{1\ {\rm erg/cm}}\right)^{-2}\left(\frac{\nu_{\rm max}}{1\ {\rm Hz}}\right)^{5}\ {\rm G}.
\end{eqnarray*}
$\tau_{\rm max}$ constitutes the optical density corresponding to the frequency $\nu_{\rm max}$ of the maximum intensity. The functions $a(p)$ and $b(p)$ are of the order 1 and can be found in Longair~(1994). The magnetic field actually present can be written as
\[B=b B_{\rm pl},\]
where the factor $b$ is dependent on the electron spectrum (cf. Sections~\ref{sec:magnetic} and~\ref{sec:maximum}). In addition, the peak brightness temperature of the source is defined as
\[T_{\max}\equiv\frac{c^{2}}{2k_{\rm B}\nu_{\rm max}^{2}}\frac{F_{\rm max}}{\pi\theta^{2}/4}.\]
We now investigate the influence of the low energy electron spectrum on the limiting brightness temperature $T_{\rm max}^{\rm c}\sim10^{12}\ {\rm K}$ for the onset of the inverse Compton catastrophe, i.e. the temperature $T_{\rm max}$ for which $P_{\rm c}/P_{\rm syn}=1$. Combining the formulae obtained so far, including second order inverse Compton losses, and choosing the power law index $p=-3$ yields
\begin{eqnarray*}
\lefteqn{\frac{P_{\rm c}}{P_{\rm syn}}=0.52\frac{\xi_{\rm geom}}{b^{2}}\left(\frac{T_{\rm max}}{10^{12}\ {\rm K}}\right)^{5}\left(\frac{\nu_{\rm c}}{1\ {\rm MHz}}\right)\cdot}\\
& & \hspace*{3.45em}\mbox{}\cdot\left[1+0.52\frac{\xi_{\rm geom}}{b^{2}}\left(\frac{T_{\rm max}}{10^{12}\ {\rm K}}\right)^{5}\left(\frac{\nu_{\rm c}}{1\ {\rm MHz}}\right)\right].
\end{eqnarray*}
If we assume that $0.3<b=B/B_{\rm pl}<3$, we see that the values of $T_{\rm max}^{\rm c}$ corresponding to $B_{\rm pl}$ and $B$ differ by a factor less than approximately $3^{2/5}\approx1.6$. Hence, the uncertainty in the magnetic field determination due to the unknown optically thick electron spectrum has a rather small impact on the limiting brightness temperature. Indeed, the uncertainty in the source geometry (i.e. in $\xi_{\rm geom}$) may be considerably more important.

\section{Conclusions}\label{sec:summary}
Based on the discussion of stationary solutions to the kinetic equations in a homogeneous and isotropic synchrotron plasma, the influence of the optically thick part of the electron spectrum on the magnetic field determination from the position of the turnover in the photon spectrum has been investigated. By employing the assumption that the optical depth equals unity at the turnover frequency it was shown that the error made by choosing some specific form for the low energy electron spectrum can be estimated to be within one order of magnitude. For power law distributions this result was corroborated by a more rigorous treatment. All results obtained are independent from the spatial extension of the source and the overall number of electrons. They show, for example, that the limiting brightness temperature is influenced to a small extent only.

Thus we have seen that that determining a magnetic field strength from the turnover of a synchrotron spectrum yields a value that is of the correct order of magnitude. Of course, in actual computations other uncertainties such as in the source geometry must also be taken into account.
\section*{Acknowledgments}
We wish to thank Klaus Beuermann and Frank Rieger for many helpful discussions. Part of this work has been supported by the Graduiertenkolleg ``Str\"{o}\-mungs\-in\-stabi\-li\-t\"{a}\-ten und Turbulenz'' of the Deutsche Forschungsgemeinschaft (DFG).

\section*{Bibliography}
Ginzburg, V.L., Syrovatskii, S.I., 1965, ARA\&A 3, 297\\[0.5ex]
Jones, T.W., 1988, ApJ 332, 678\\[0.5ex]
Kellermann, K.I., 1974, in Verschuur, G.L., Kellermann, K.I., eds, Galactic and\\
\hspace*{1.7em}Extra-Galactic Radio Astronomy. Springer, New York, p.~320\\[0.5ex]
Longair, M.S., 1994, High energy astrophysics (Volume 2), Cambridge Univer-\\
\hspace*{1.7em}sity Press, Cambridge\\[0.5ex]
McCray, R., 1969, ApJ 156, 329\\[0.5ex]
Melrose, D.B., 1980, Plasma Astrophysics (Volume 1), Gordon and Breach, New\\
\hspace*{1.7em}York\\[0.5ex]
Norman, C.A., ter Haar, D., 1974, Phys. Rep. 17, 307\\[0.5ex]
Pacholczyk, A.G., 1970, Radio Astrophysics, W. H. Freeman and Company, San\\
\hspace*{1.7em}Francisco\\[0.5ex]
Rybicki, G.B., Lightman, A.P., 1979, Radiative Processes in Astrophysics,\mbox{Wiley},\\
\hspace*{1.7em}New York\\[0.5ex]
Westfold, K.C., 1959, ApJ 130, 241
\end{document}